\documentstyle[12pt]{article}
\begin{document}

\newcommand{\be}{\begin{equation}}
\newcommand{\ee}{\end{equation}}
\newcommand{\<}{\langle}
\renewcommand{\>}{\rangle}
\def\reff#1{(\protect\ref{#1})}
\newcommand{\C}{{\cal C}}

\title{Random walks with long-range self-repulsion on proper time}
\author{
  \\
  { Sergio Caracciolo}              \\[-0.2cm]
  {\small\it Scuola Normale Superiore and INFN -- Sezione di Pisa}
  \\[-0.2cm]
  {\small\it Piazza dei Cavalieri}        \\[-0.2cm]
  {\small\it Pisa 56100, ITALIA}          \\[-0.2cm]
  \\
  { Giorgio Parisi}                  \\[-0.2cm]
  {\small\it Dipartimento di Fisica and INFN -- Sezione di Roma-Tor Vergata}
       \\[-0.2cm]
  {\small\it Universit\`{a}  di Roma II}         \\[-0.2cm]
  {\small\it Via E.~Carnevale}          \\[-0.2cm]
  {\small\it Roma 00173, ITALIA}      \\[-0.2cm]
  \\  \and
  { Andrea Pelissetto}              \\[-0.2cm]
  {\small\it Dipartimento di Fisica}       \\[-0.2cm]
  {\small\it  Universit\`{a}  di Pisa and INFN -- Sezione di Pisa}
  \\[-0.2cm]
  {\small\it Piazza Torricelli}          \\[-0.2cm]
  {\small\it Pisa 56100, ITALIA}          \\[-0.2cm]
  {\protect\makebox[5in]{\quad}} 
  \\
}

\date{}
\maketitle
\thispagestyle{empty}   

\vspace{0.5 cm}

\begin{abstract}
We introduce a model of self-repelling random walks
where the short-range interaction between two elements of the chain decreases
as
a power of the difference in proper time.
Analytic results on the exponent $\nu$ are obtained. They are in good agreement
with Monte Carlo simulations in two dimensions. A numerical study of the
scaling functions and of the efficiency of the algorithm is also presented.
\end{abstract}

\clearpage

\section{Introduction}

Self-avoiding walks are a fascinating subject.
Although they are much simpler than an Ising or
Heisenberg spin system, they behave quite
similarly in the critical region where the
coherence length is much larger than the lattice
spacing. There are deep reasons for this
similarity. Indeed, using Symanzik
representation~\cite{Symanzik,BFS}, the free energy of a spin system
may be written as sum over partition
functions of self-avoiding walks. Moreover
the self-avoiding
walks are exactly the limit of an $O(n)$-invariant
spin system for $n$ going to zero~\cite{DeGennes,desCl75,Daoud,Emery}.

A long time ago Flory~\cite{Flory}, using simple minded
approximations, found that the critical
exponent $\nu$ in dimension $D$ less than or equal
to 4,  was given by the simple expression
\be
\nu = 2/(3+D) \label{1}
\ee
This result is quite puzzling; it is exact for
$D=$1, 2 and 4, but it is definitely wrong in
$D=3$~\cite{Le-Guillou,Guttmann,Madras,CFP}. Moreover in $4 - \epsilon$
dimensions the exact results and Flory's one differ at first order in
$\epsilon$~\cite{DeGennes}.
The Flory approximation was strongly criticized by
des~Cloiseaux~\cite{DeCloiseaux,desCl-book}, but he found the puzzling result
that a more accurate variational approximation
leads in three dimensions to the rather bad
value of $\nu=2/3$.

Quite recently it was found that des~Cloiseaux' method can be extended to the
case of polymers with long-range repulsion~\cite{BMPY}. Also in
this case Flory's and des~Cloiseaux' approaches
give  different results, however here it is
possible to give very strong arguments
suggesting that the variational approach of
des~Cloiseaux gives the exact result in the case
where the space dimension becomes infinite.

The putative exact infinite-dimensional
results can be extrapolated at finite
dimensions using arguments based on the
renormalization group and performing the
approximation directly at finite dimension.
One finally finds that for an
interaction which decreases with the distance as a power law
with exponent $\lambda$, $\nu$
 is given by the generalization of
des~Cloiseaux' formula~\cite{BMPY}
\be
\nu = \cases{1 & for $\lambda \leq 2$ \cr
      2/\lambda & for $2 \leq \lambda \leq 4$ \cr
      1/2 & for $\lambda \geq 4$}
\ee
with logarithmic corrections when $\lambda=2,4$.

Some of these predictions have been
numerically tested~\cite{MP}.

A very different case concerns a potential which is
short-range in space but in which the interaction among
different elements of the polymer
decreases as a power of the distance along the
chain. An interesting theoretical analysis of
this model can be done and the predictions can be
tested using numerical
methods due to the short-range nature of the
Hamiltonian.

After this introduction we present the
model in section 2. An
approximate computation of the exponent $\nu$
is presented in section 3. The algorithm we use
for numerical simulations is described in
section 4, while the results we obtain are presented in section 5. Some
conclusions are presented in section 6.

\section{The model}

Let $\C$ be the ensemble of random walks
on the hypercubic lattice ${\bf Z}^D$ of $N$
steps, starting from the origin. Thus, if $\omega\in \C$
\be
\omega = \{\, \vec{\omega}_0, \vec{\omega}_1, \vec{\omega}_2, \ldots,
\vec{\omega}_N \}
\ee
with $\vec{\omega}_0=\vec{0}$ the origin of the lattice,
$\vec{\omega}_i\in{\bf
Z}^D$ the location at {\em time} $i$ of the walk and $|\vec{\omega}_i -
\vec{\omega}_{i-1}|=1$ for $i=1,\ldots,N$.

We consider an interaction of the form
\be
H_I[\omega] = {1\over 2}\, g\,N^\delta \sum_{{i,j=0}\atop i \neq j}^N
{\delta_{\vec{\omega}_i , \vec{\omega}_j}\over |i-j|^\lambda} \label{HI}
\ee
This means that in the ensemble average $\langle\cdot\rangle$ in $\C$ we
associate to each walk the statistical weight
\be
m[\omega] = {e^{-H_I[\omega]}\over Z_N[g] }
\ee
where
\be
Z_N[g] = \sum_{\omega\in\C} e^{-H_I[\omega]}
\ee
To study the conformation of the walks in this ensemble we shall consider
the square end-to-end distance
\be R_e^2[\omega] \; \equiv \; \vec{\omega}_N^2 \ee
and the square radius of gyration
\be
 R_g^2 [ \omega ]   \; \equiv \;   {1 \over N+1}  \sum_{i=0}^N
        \left( \vec{\omega}_i - {1 \over N+1} \sum_{j=0}^N \vec{\omega}_j
\right)^2 \ee
Both of them are believed to have the asymptotic behaviour
\be
\< R^2 \> \sim  N^{{2} \nu} \label{nu}
\ee
as $N \to \infty$, with the same critical exponent $\nu$.

We shall also consider the universal ratio
\be
A \equiv {\< R_g^2 \>\over \< R_e^2 \>}
\ee
For the ORW we have
$A=1/6\sim 0.16667$, while for the SAW $A$ is known only numerically and
depends on the dimension of the embedding space. In two dimensions
the most precise estimate is obtained by a Monte Carlo
simulation~\cite{CPS} whose result is $A= 0.14026 \pm 0.00011$, where the error
bar is $95\%$ level of confidence.

\section{Heuristic analysis}

In this Section we want to derive an estimate of the critical exponent $\nu$
with a variational approach.
Let us firstly discuss the case $\delta=0$~\cite{BMPY}.

Following Des Cloiseaux~\cite{DeCloiseaux} one considers random rings
instead of random walks,
as this fact does not change the value of $\nu$ and simplifies the calculations
by permitting Fourier analysis along the chain. In addition it is simpler to
work in continuum space rather than on a lattice. In this case one must somehow
regularize the $\delta$-function  appearing in \reff{HI}. We will thus consider
as equilibrium  probability measure for the model
\be
dm(\omega) = {1\over Z} \exp (- H)\,
d^D\vec{\omega}_1
d^D\vec{\omega}_2 \ldots d^D\vec{\omega}_N
\ee
where the Hamiltonian is given by
\be
H =
{1\over 2} \sum_{i=0}^N (\vec{\omega}_i -
\vec{\omega}_{i-1})^2 + {1\over 2}\, g\, \sum_{{i,j=0}\atop i \neq j}^N
{V\left[\left(\vec{\omega}_i - \vec{\omega}_j\right)^2\right]
\over |i-j|^\lambda}
\ee
Here $V(x^2)$ can be any arbitrary short-range potential and for definiteness
we will assume
\be
V(x^2) = \exp\left(-{x^2\over 2 a}\right)
\ee
The mean field approximation is based on a variational approach
with a Gaussian trial measure~\cite{DeCloiseaux,BMPY}
\be
dm_0(\omega) = {1\over Z_0} \exp (- H_0)\,
d^D\vec{\omega}_1
d^D\vec{\omega}_2 \ldots d^D\vec{\omega}_N
\ee
with
\be
H_0 = {1\over 2} \sum_{i,j=0}^N  G_{ij}^{-1}
\,\vec{\omega}_i \cdot \vec{\omega}_j
\ee
and
\be
Z_0 = (2  \pi)^{(N+1)D/2} (\det G)^{D/2}
\ee
The function $G_{ij}$ is determined by minimizing the functional
\be
F[G] = \langle H - H_0\>_0 \, -\, \log Z_0
\ee
where $\<\,\,\>_0$ denotes the expectation value with respect to
$dm_0$.

Because of invariance under translations along the chain we have
$G_{ij} = G(i-j)$ and by symmetry  $G(n) = G(-n)$.
It is also convenient to introduce the Fourier transform
\be
\tilde{G} (p) = \sum_{\tau=0}^{N-1} G(\tau) e^{ip\tau}
\ee
where $p= 0, 2\pi/N, \ldots, 2\pi(N-1)/N$, and  $\tilde{G} (p)$ is real and
positive due to the positive definiteness of  $G_{ij}$.

The functional $F$ can be computed as
\begin{eqnarray}
{F[G]\over N} &=& D \int_{-\pi}^\pi {dp\over 2\pi} \tilde{G}(p) \,
\left(1-\cos p\right) - {D\over 2}\, \left[1+ \log (2 \pi) \right] - {D\over 2}
\int_{-\pi}^\pi {dp\over 2\pi} \log \tilde{G}(p) \nonumber\\
&& +  {g\over 2}\sum_{\tau=1}^\infty {1\over \tau^\lambda} \left[
1 + {2\over a} \int_{-\pi}^\pi {dp\over 2\pi} \tilde{G}(p)
\left(1-\cos p\tau\right)  \right]^{-{D\over 2}}
\end{eqnarray}
where we have replaced sums over $p$ with the corresponding integrals as we
are interested in the regime of very large $N$.

The minimization condition becomes
\begin{eqnarray}
\lefteqn{{1\over \tilde{G}(p)} = 2 \left(1-\cos p\right)} \label{G-minimo}\\
&&- {g\over  a} \sum_{\tau=1}^\infty {1\over \tau^\lambda}
 \left(1-\cos p\tau\right) \left[
1 + {2\over a} \int_{-\pi}^\pi {dp'\over 2\pi} \tilde{G}(p') \left(1-\cos
p'\tau\right) \right]^{-{D+2\over 2}} \nonumber
\end{eqnarray}
The exponent $\nu$ can be extracted, as easily seen, from the low-momentum
behaviour of $\tilde{G} (p)$ as
\be \tilde{G}(p) \sim {1\over |p|^{2\nu +1}} \ee
The analysis of \reff{G-minimo} is quite subtle and can be done following the
original  paper by Des Cloiseaux~\cite{DeCloiseaux} and predicts $\nu=1/2$ for
$D>4$
while for $2\leq D\leq 4$ gives the following result~\cite{BMPY}
\be
\nu_{MF}(\lambda,D) = \cases{1/2 & for $\lambda > {1\over2} (4-D)$ \cr
      {1\over D} (2-\lambda) & for $\lambda < {1\over2} (4-D) $}\label{22}
\ee
For $\lambda=0$, $D=2$ additional logarithmic corrections appear as
\be
\langle R^2_g\rangle \sim N^2/ \log N \ee
and analogously for $\lambda =  {1\over2} (4-D)$ we get
\be
\langle R^2_g\rangle \sim N (\log N)^{2/D} \ee
Thus we obtain an ORW for $\lambda$ large enough. This has
indeed to be expected, as we have the trivial rigorous bound
\be
H_I (\omega) < C N^{2-\lambda} \ee
showing that at least for $\lambda>2$ the model is an ORW.

On the other hand the mean-field result $\nu_{MF}$ is definitely wrong
for $\lambda=0$, $D\leq4$, which corresponds to the SAW
limit~\cite{Edwards,DeCloiseaux}.
In two and three dimensions $\nu_{SAW}$
is equal respectively to 3/4 and approximately 3/5 (see \reff{1}) while
$\nu_{MF}(0)$
corresponds to  1 and 2/3. For $D=4$ the mean-field approach correctly predicts
logarithmic  corrections to the random-walk behaviour, but the power of the
logarithm
is higher than that obtained using the renormalization group which
predicts~\cite{RGlog}
\be
\langle R^2_g\rangle \sim N (\log N)^{1/4} \ee
Of course the variational approach will overestimate $\nu$ also for
small $\lambda$. We do not have any theoretical control of this regime.
However as we will discuss in Section 5, our numerical results are in
reasonable agreement with the following conjecture for $\nu(\lambda,D)$
\be
\nu(\lambda,D) = \min \left(\nu_{SAW}, \nu_{MF} (\lambda,D) \right) \label{25}
\ee
Let us finally notice that for large $N$, the mean-field equation
\reff{G-minimo}
has a general scaling solution. Indeed in the limit of small $p$, which
corresponds to large $N$ we can rewrite \reff{G-minimo} as
\begin{eqnarray}
\lefteqn{{1\over \tilde{G}(p)} =  p^2 } \label{G-minimo-scaling}\\
&&- g\,{1\over 2 a} p^2  \int_{1}^\infty d\tau  \tau^{2-\lambda}
  \left[
 {2\over a} \int_{-\pi}^\pi {dp'\over 2\pi} \tilde{G}(p') \left(1-\cos
p'\tau\right) \right]^{-{D+2\over 2}} \nonumber
\end{eqnarray}
It is easy to check that the general solution has the form
\be
\tilde{G} (p) = {1\over p^2} \hat{f}_\lambda(g\, p^{\lambda - 2 + D/2}) \ee
which gives
\be
 \< R_g^2 \> =  N \hat{f_\lambda}(g\, N^{2-\lambda - D/2}) \label{28}
\ee
Let us now discuss  the case $\delta\neq 0$.
For simplicity we shall set since now on $D=2$, the generalization to arbitrary
$D$
being trivial.

Let us firstly consider the case in which $\delta<0$, namely the case in which
we are weakening the coupling constant when approaching the asymptotic limit.
One can think at the model in the intermediate region in a perturbative
expansion around the ORW, that is around $g=0$. To evaluate the dimension of
the
coupling constant, remark that $N^{2 - \nu D}$ is the asymptotic behaviour for
large
$N$ of the average number of intersections of two walks of $N$ steps and
Hausdorff
dimension $1/\nu$ on a lattice of dimension $D$. Then
\be [g] = - [N] (\delta -\lambda +2 -\nu_{ORW} D) = 2 + 2\,\delta - 2\,\lambda
\ee
and
\be [N] = - {1\over \nu_{ORW}} = - 2\ee
Thus in the scaling region we
expect that for $g N^\delta$ small
\be
 \< R_g^2 \> = N\, F_\lambda(g, N) = N f_\lambda(g\, N^{1+\delta-\lambda})
\ee
Let us notice that for $\delta=0$ this expression coincides with the scaling
formula \reff{28}.
For $g=0$ the model is an ORW and thus we get $f_\lambda(0)
\neq 0$. It follows that, when $\delta<\lambda-1$ $ \< R_g^2 \>$ scales as $N$.
On the other hand when
$\lambda-1\leq\delta<0$ the argument of the scaling function $f_\lambda$
goes to infinity. Assuming in this limit $f_\lambda (x) \sim x^\beta$
we get
\be
 \< R_g^2 \> \sim  N^{1 + \beta ( 1 + \delta - \lambda)}
\ee
The value of $\beta$ is computed using the conjectured value for $\nu$
when $\delta=0$. In this way we obtain that $f_\lambda (x)$ scales for large
$x$
as
\be
f_\lambda(x) \sim \cases{
x^{2 \nu_{SAW} - 1\over 1 -\lambda} \sim x^{1\over 2(1 -\lambda)} & when $0
                    \leq\lambda\leq 1/2$\cr
x^{2 \nu(\lambda) - 1\over 1 -\lambda} \sim x & when $1/2
                    \leq\lambda\leq 1$\cr }
\ee
Then we obtain the prediction
\be
 \< R_g^2 \> \sim  \cases{
N^{{3\over2}+{\delta\over 2(1-\lambda)}}& when $0
                    \leq\lambda\leq 1/2$, $\delta\leq\lambda-1$\cr
N^{2 + \delta-\lambda}& when $1/2
                    \leq\lambda\leq 1$, $\delta\leq\lambda-1$\cr
N   & when $\lambda\geq 1$ and $\lambda<1$, $\delta > \lambda-1$\cr }
\label{scaling-n}
\ee
In the opposite situation ($\delta>0$) the coupling constant goes to
infinity while approaching the asymptotic limit and we can expect that
in this case the scaling behaviour is controlled by the SAW fixed point.
Then
\be [N] = - {1\over \nu_{SAW}} = - {4\over 3}\ee
and
\be [g] = - [N] (\delta - \lambda + 2 - \nu_{SAW} D)
\ee
so that we expect that for $g N^\delta$ large
\be
 \< R_g^2 \> = N^{2\nu_{SAW}}\, \bar{F}_\lambda(g, N) = N^{3\over 2}
\bar{f}_\lambda(g\, N^{{1\over 2}+\delta-\lambda}) \label{39}
\ee
Since for $g$ going to infinity at fixed $N$ we expect the model to describe a
SAW, $\bar{f}_\lambda (x) $ must converge to a constant for large $x$.
It immediately follows that for $\delta>\lambda - {1\over 2}$ the behaviour is
that of a SAW.
On the other hand, when $\delta<\lambda-1/2$ and $\lambda > 1/2$,
the argument of the scaling function goes to zero when $N$ goes to infinity.
As for $g=0$ we have an ORW, we must have
$\bar{f}_\lambda (0) = 0 $. Then, assuming for small $x$ the scaling form
$\bar{f}_\lambda (x) \sim x^\beta $ we get
\be
 \< R_g^2 \> \sim  N^{3/2 + \beta ( 1/2 + \delta - \lambda)}
\ee
This exponent $\beta$ is computed by requiring that for $\delta = 0$ this
formula
reproduces the result \reff{25}. We thus get that the scaling
function $\bar{f}_\lambda$ must scale for small argument as
\be
\bar{f}_\lambda(x) \sim \cases{
x^{4 (\nu(\lambda) - \nu_{SAW})\over 1 -2 \lambda} \sim x & when $1/2
               \leq\lambda\leq 1$ \cr
x^{4 (\nu_{ORW} - \nu_{SAW})\over 1 -2 \lambda} \sim x^{1\over 2\lambda-1} &
when
               $\lambda\geq 1$ \cr}
\ee
Then for positive $\delta$ we get
\be
 \< R_g^2 \> \sim \cases{
 N^{3/2} & when $0\leq\lambda\leq 1/2$ and $\lambda>1/2$, $\delta>\lambda-1/2$
\cr
 N^{2 + \delta-\lambda }& when $1/2\leq\lambda\leq 1$, $\delta\leq\lambda-1/2$
\cr
 N^{1+{\delta\over 2\lambda-1}}& when $\lambda\geq 1$
$\delta\leq\lambda-1/2$\cr}
\label{scaling-p}
\ee
Let us notice that all these scaling arguments do not take into
account logarithmic corrections which we expect to be present for those
values of $\delta$ where there is the transition to the purely SAW or
ORW behaviour.

\section{The algorithm}

We have simulated the model by using the so-called {\em pivot}
algorithm~\cite{Lal,Mac,MS}, which is known  to be  extremely  efficient for
the simulation of SAWs with fixed number of steps and free end-points as the
computer time necessary to produce an independent walk is of the order of the
number of steps in the walk, which is also the best possible behaviour because
this is the order of time necessary simply to write down all the steps.

The algorithm is defined as follows~\cite{MS}. Choose at random a point along
the walk (the {\em pivot}), but not the first or the last one. Let it be
the $k$-th point $\vec{\omega}_k$, with $0<k<N$. Then choose at random an
element $g$ in the symmetry group of the lattice and propose a new walk
$\omega'$ defined by
\be
\vec{\omega}_k' = \cases{
 \vec{\omega}_i & for $0\leq i \leq k$ \cr
 \vec{\omega}_k + g(\vec{\omega}_i- \vec{\omega}_k) & for $k+1\leq i \leq N$
  \cr
}
\ee
The new walk is accepted according to a Metropolis test in order to
generate the desired statistical ensemble.

For the limiting case of ORW and SAW it is known~\cite{MS} that for  a global
observable $A$ the integrated autocorrelation time
$\tau_{int,A}$ scales for large number of steps according to
\be
\tau_{int,A} \sim N^p
\ee
where (the exponent $p$ should be the same for all global variables)
\be
p = \cases{ 0 & for ORW \cr 0.194 \pm 0.002 & for SAW \cr}
\label{scaling-dinamico}
\ee
For our models, as it can be seen from Table~\ref{tempi}, where we report the
integrated autocorrelation time for the end-to-end distance, we obtain similar
results. When the values of the parameters are such that the walks are in the
same universality class of ORW or SAW, the dynamical behaviour is compatible
with~\reff{scaling-dinamico}. In the intermediate cases for which
$1/2\leq\nu\leq 3/4$   the dynamic critical
exponent $p$
ranges  correspondingly within the interval $[0,0.19]$, and we find that
it is in reasonable agreement with  a linear interpolation of the form
\be
p_\nu = 0.39\, (2\,\nu -1)
\ee
Let us now come to the computational complexity.
In the practical implementation we used a hash-table with linear probing in
order to check for the self-intersections of the walk.
In the case of SAWs Madras and Sokal~\cite{MS} showed that it is particularly
convenient to insert the points in the hash-table starting from the pivot
point and working outward (and of course stopping as soon as an intersection is
detected). In this way the mean work per move turns out to be of order
$N^{1-p}$, thus smaller than the work done by inserting the points without a
special order which is of order $N$.
We used in our case a similar trick. Indeed, chosen a random number $r$
uniformly distributed in the unit interval, according to the Metropolis
prescription, the proposed walk $\omega'$ is accepted if
\be
H[\omega'] \leq  S \equiv H[\omega] - \ln r
\ee
Our implementation works as follows: after the choice of the pivot point and
of the transformation $g$  we choose the random number $r$ and compute the
quantity $S$. Then we begin to construct the new walk and we insert the points
in the hash-table starting from the pivot point and working outward. Whenever
we find an intersection we sum to the energy of the new walk the contribution
from that intersection, but we stop if the accumulated value is already larger
than $S$, because the proposed walk has to be rejected~\footnote{
This procedure works because the interaction is repulsive. For
attractive interactions, as for instance for SAWs or trails at the
$\theta$-point, the whole new walk must be defined and therefore the mean work
per pivot-move will scale as the number of step $N$.}.
In the limiting case of very strong repulsion, when the universality class is
the same of SAW, this implementation works exactly as the original one devised
by Madras and Sokal. In the opposite limiting case, the ORW-one, practically
all proposed walks will be accepted and thus no improvement can be expected.
In the general situation we expect (although we didn't check) that the
computational work scales as $N^{(1-p)}$ where $p$ is the dynamic critical
exponent for global observables and thus, in all cases, that the computer
time necessary to produce a statistically independent measurement is of
order $N$.
\begin{table}
\centering\begin{tabular}{|cl|ccccc|}
\hline\hline
\multicolumn{2}{|c|}{$\lambda$} & 0.00 & 0.00 & 0.00 & 0.00 & 0.25\\ \hline
\multicolumn{2}{|c|}{$\delta$} & $-1.00$ & $-0.75$ & $-0.50$ & $-0.25$ & 0.00\\
\hline
\multicolumn{2}{|c|}{$g$} & 1.00 & 2.00 & 3.00 & 4.00 & 1.00\\ \hline
$N=$ & $100$  & 6.27(5) & 6.22(5)  & 7.00(6)   & 8.07(8)   & 7.48(7) \\
$N=$ & $200$  & 6.31(8) & 6.24(7)  & 7.76(10)  & 9.07(13)  & 8.53(12) \\
$N=$ & $500$  & 6.54(8) & 6.44(7)  & 7.99(11)  & 10.77(17) & 9.87(15) \\
$N=$ & $1000$ & 6.53(8) & 6.61(8)  & 8.62(12)  & 11.63(20) & 11.19(18) \\
$N=$ & $2000$ & 6.51(8) & 6.54(8)  & 8.74(12)  & 13.24(23) & 13.14(23) \\
$N=$ & $4000$ & 6.40(8) & 6.64(9)  & 9.59(14)  & 14.71(27) & 15.37(29) \\
\hline
\multicolumn{2}{|c|}{$p_\nu$}& 0.00 & 0.05 & 0.09 & 0.14 & 0.19 \\
\hline
\multicolumn{2}{|c|}{$p$} & 0.00(1) & 0.01(1) & 0.08(1) & 0.15(1) & 0.20(1) \\
\hline \hline
\end{tabular}
\caption{Integrated autocorrelation times for the end-to-end distance for
selected values of the parameters $\lambda$, $\delta$ and $g$ and of the number
of steps $N$ of the walks. $p_\nu$ is the value obtained through a linear
interpolation for the dynamic critical exponent, and $p$ is the estimate based
on a fit from our data.
         }
\label{tempi}
\end{table}

\section{Numerical results}

In order to test the ideas presented in Sect. 3 we have performed an extensive
Monte
Carlo simulation on a square lattice on walks of lengths ranging from 100 to
8000.
The total CPU time for these runs was roughly 2000 hours of a VAX 6000-520.

In the numerical simulation the first problem one has to deal with is the
initialization, that is,  for each value of $\lambda$, $\delta$, $g$ and
$N$ one has to generate a starting walk. This was done in two different ways
according
to the value of $N$. When $N$ was less than or equal to 2000 we generated a SAW
using
a dimerization routine~\cite{Suzuki,Alex1,Alex2,MS}. When instead $N$ = 4000,
8000,
as the dimerization routine is too costly
(the computer time needed to generate a walk scales as $\tau\sim N^{a \log_2 N
+ b}$,
with $a\approx 0.17$ and $b\approx 0.72$ in two dimensions),
we used
the scanning method~\cite{Meirovitch} with scanning parameter equal to
3~\footnote{ The only reason why we used the dimerization routine for the
low-$N$ runs was its availability at the time of the runs. In retrospect it
would have probably been better to use the scanning program in all cases.}.
None of these two methods generates a random sample of walks with the correct
equilibrium distribution, although the first  is exact in the SAW-limit and the
second in the ORW-one. It was thus necessary to run a certain number of
thermalization iterations before measuring. As the convergence to equilibrium
of a Markov chain is controlled by the exponential autocorrelation time
$\tau_{exp}$ it is necessary to run a few $\tau_{exp}$ iterations to
reach equilibrium. For the pivot algorithm $\tau_{exp}$ is proportional
to $N$. For this
reason  we ran approximately $10N$ pivot iterations for thermalization before
measuring. For each value of $\lambda$, $\delta$, $g$ and $N$ we have then
performed
$10^6$ iterations, except when $N=100$ in which case the runs consist of
$2\,10^6$
iterations. The integrated autocorrelation times for the squared end-to-end
distance
and  for the square radius of gyration range from 3 to 20 and from 5 to 60
respectively. We have reported a few of them in Table~1.

In Table~\ref{table1} and Table~\ref{table3}
we report the results of our runs
for $\< R_e^2 \>$  for different values of the parameters,
respectively for negative and positive values of $\delta$.

We have performed least-squares regressions on these data in order to determine
the
critical exponent $\nu$. We fit $\< R^2_e\>$ to the Ansatz $aN^{2\nu}$ by
performing a weighted least-squares regression of its logarithm against $\log
N$,
using the a priori error bars on the raw data points to determine both the
weights and the error bars. In order to control the systematic error
due to corrections to scaling we have done various fits in which the data
points
with lowest $N$ were discarded. The results of these fits are reported in
Table~4 and Table~5 for various $N_{cut}$, where $N_{cut}$ is the minimum $N$
included in the fit. In many cases it is evident a systematic drift of the
estimated exponent with $N_{cut}$, an indication of strong corrections to the
scaling. This effect is however strongly dependent on the value of $g$.
When the expected value of $\nu$ is different from $1/2$ and $3/4$ one observes
that
for $g$ small the estimated value of $\nu$ increases with $N_{cut}$, while for
$g$
large the estimate decreases. For the intermediate values of
$g$ there is a flatness region where $\nu$ remains
approximately constant meaning that for these values the $g$-dependent
corrections
are  small compared to our statistical error. For every $\lambda$ and $\delta$
we have made various runs for different values of $g$ in order to find the
flat region  were the corrections are small enough, in order to obtain
estimates of
$\nu$ with a  smaller systematic error. Our final estimates are in reasonable
agreement with the  proposed value of $\nu$, the discrepancy being less than a
few
per cent. The worst cases are for $\lambda=1$ and $\lambda=0.5$. For instance
for $\delta=1/3$, $\lambda=1$ our data suggest $\nu \approx 0.68,0.69$ while
the
expected value is  $\nu=2/3\approx 0.667$. Let us notice however that for
$\lambda=1$ and $\lambda=0.5$ we expect the presence of logarithmic
corrections,
i.e. a behaviour of the form
\be
\< R^2_e \>_N \, =\, a N^{2\nu} \log^\beta N (1 + \, O(1/\log N,\, \log\log
N/\log
N))
\ee
The presence of logarithms makes the analysis very difficult. First of all
in this case the convergence to the asymptotic regime is extremely
slow. Moreover the presence of the term $\log^\beta N$ makes impossible
an evaluation of $\nu$. Indeed as we use data with $200\leq N\leq 8000$,
$\log^\beta N$ behaves, as far as the fit is concerned, approximately as
$N^{\approx 0.3\beta}$. Thus in a pure power-law fit one really measures $2\nu
+
0.3\beta$. Since we don't have any theoretical knowledge of $\beta$ it is thus
impossible  to draw any definite conclusion.

To better understand the validity of our
conjecture \reff{22} we have made two runs with higher statistics at
$\lambda=0.75$ and $\lambda=0.25$ with $\delta=0$. Each data point corresponds
here to $9\times 10^6$ iterations and in order to avoid any initialization
bias we have discarded the first 100 $N$ iterations. The results are
reported in Table~\ref{tabp}. A good agreement is seen although a systematic
trend is visible in both cases.
  \begin{table}
\centering\begin{tabular}{|c|r|l|r|l|}
\hline\hline
 & \multicolumn{1}{c|}{$R^2_e$} & \multicolumn{1}{c|}{$\nu$} &
\multicolumn{1}{c|}{$R^2_e$} & \multicolumn{1}{c|}{$\nu$}\\ \hline
$N$&\multicolumn{2}{c|}{$\lambda=0.25$}& \multicolumn{2}{c|}{$\lambda=0.75$}\\
\hline
500\vphantom{0}  & 4862(10)   & 0.7433(11) & 1874(4)     & 0.6238(6) \\
1000 & 13554(31)  & 0.7447(17) & 4449(10)    & 0.6239(8) \\
2000 & 37927(91)  & 0.7460(27) & 10556(25)   & 0.6242(13) \\
4000 & 106704(277)&            & 25073(60)   &  \\
8000 & 300005(874)& & 59567(155)   &  \\ \hline
$\nu(\lambda)$ & & 0.7500 & & 0.6250 \\ \hline\hline
\end{tabular}
\caption{Results for the runs with higher statistics and comparison with the
expected value for the exponent $\nu$. In both cases $\delta=0$ and $g=1$.
        }
\label{tabp}
\end{table}

We have then checked if $\< R^2_e\>$ obeys the scaling laws \reff{28} and
\reff{39}.
In Fig.~\ref{Fig1} we plot our estimates of $\< R_e^2 \>/N$ versus
$N^{1+\delta-\lambda}$ for $\delta<0$ and $\lambda=0.00$, 0.25, 0.50, 0.90 with
$N>200$. The agreement seems quite good.
However a more detailed examination of the scaling plot shows that the data
points
do not belong to a unique curve within error bars. Indeed one sees that
different
runs with the same values of $g$ and $\delta$ belong to distinct curves which
approach each other only when $N$ becomes large. This fact has to be expected.
Indeed, as the analysis of the exponent $\nu$ shows, for the values of $N$
which we
are considering the corrections to scaling are still large and thus we expect
analogously large violations to the scaling behaviour given by \reff{39}.
\begin{figure}[p] \vspace*{10cm} \hspace*{-1cm} \special{illustration
figura_parisi1}  \caption{Scaling function for the end to end distance in the
case
$\delta<0$. Small pentagons, rectangles, triangles and diamonds correspond
respectively to the values $\lambda=0.90$, 0.50, 0.25 and 0.
 } \label{Fig1}
\end{figure}
Analogously in Fig.~\ref{Fig3} we plot our estimates $\< R_e^2 \>/N^{3/2}$
versus
$N^{1/2+\delta-\lambda}$ for $\delta<0$ and $\lambda=0.75$, 0.90 and $N>200$.
Here again the agreement is only approximate.
The situation is even worse
for $\lambda=1.00$ and $\delta>0$: in this case the points are scattered and no
scaling can be observed. This can be explained by the presence of logarithmic
terms,
which break the scaling laws and make the approach to the asymptotic
regime extremely slow.
\begin{figure}[p] \vspace*{10cm} \hspace*{-1cm}
\special{illustration figura_parisi3}  \caption{Scaling function for the end to
end
distance in the case $\delta>0$. Small pentagons and triangles correspond
respectively to the values $\lambda=0.90$ and 0.75~.
 } \label{Fig3}
\end{figure}
As a final check we have studied the behaviour of the universal ratio $A$,
which,
using the scaling laws \reff{28} and \reff{39} must have the form
\be
A = h_\lambda (g N^{1+\delta-\lambda})
\ee
for $\delta<0$ and
\be
A = \overline h_\lambda (g N^{1/2+\delta-\lambda})
\ee
for $\delta>0$.
Moreover for $x\to0$ both functions must converge to the ORW-value 1/6.
In Fig.~\ref{Fig2}  and \ref{Fig4} we present the scaling plots for A in the
two
cases.  The agreement is reasonable, although a closer inspection shows again
the
presence of systematic deviations.
\begin{figure}[p] \vspace*{10cm} \hspace*{-1cm}
\special{illustration figura_parisi2}
\caption{Scaling function for the universal ratio $A$ in the case $\delta<0$.
Small pentagons, rectangles, triangles and diamonds correspond respectively to
the
values $\lambda=0.90$, 0.50, 0.25 and 0.
 } \label{Fig2}
\end{figure}
\begin{figure}[p] \vspace*{10cm} \hspace*{-1cm}
\special{illustration figura_parisi4}
\caption{Scaling function for the universal ratio $A$ in the case $\delta>0$.
Small pentagons and triangles correspond respectively to the
values $\lambda=0.90$ and 0.75~.
 } \label{Fig4}
\end{figure}

\section{Conclusions}

In conclusion the results of our simulations are in good agreement with the
theoretical arguments of Sect. 3, suggesting that our conjectured value for
$\nu$, if not exact, is certainly a very good approximation in the
range of parameters we have examined.

A Flory argument would predict for the exponent $\nu$
\be
\nu_F = {3 - \lambda + \delta\over 2 + D}
\ee
It is interesting to remark that such a value is a good approximation,
according
to our analysis, only for $\lambda=\delta=0$. The deep reason for such an
agreement escapes  us.

\section*{Acknowledgments}

S.C. and A.P.  wish to thank Alan D.~Sokal for useful discussions.

\clearpage

\begin{table}
 \protect\footnotesize
\hspace*{-2cm}\begin{tabular}{|cc|ccccccc|}
\hline\hline
$\delta$ & g & 100 & 200 & 500 & 1000 & 2000 & 4000 & 8000\\ \hline
& & \multicolumn{7}{||c||}{$\lambda=0$}\\ \hline
$-$1 & 10. & 207(1)& 414(2)& 1045(6)& 2071(12)&
              4169(24)& 8284(48) & \\
\hline
$-$3/4 & 2.0 & 175(1)& 369(2)& 1016(5)& 2187(12)&
              4702(26)& 10158(58) & \\
$-$3/4 & 5.0 & 249(1)& 541(3)& 1504(8)& 3268(18)&
              7084(38)& 15342(86) & \\
$-$3/4 & 8.0 & 307(1)& 669(3)& 1874(10)& 4062(122)&
              8876(50) & 19390(108) & \\
\hline
$-$1/2 & 1.0 & 207(1)& 477(3)& 1460(8)& 3475(19)&
              8148(46)& 19273(108) & \\
$-$1/2 & 3.0 & 332(1)& 786(4)& 2480(13)& 5871(33)&
              14000(80) & 33389(198) & \\
$-$1/2 & 5.0 & 417(1)& 1000(5)& 3164(18)& 7624(44)&
              18227(110) & 42895(264) & \\
\hline
$-$1/4 & 1.0 & 341(1)& 876(5)& 3071(17)& 7980(46) &
              20780(124) & 54207(350) & 140178(932)\\
$-$1/4 & 2.0 & 462(2)& 1203(7)& 4285(25)& 11087(68)&
              28817(188) & 75441(518)  & \\
$-$1/4 & 4.0 & 612(2)& 1621(9)& 5815(36)& 15225(100) &
              39867(274)& 104450(754) & \\
\hline
& & \multicolumn{7}{||c||}{$\lambda=0.25$}\\ \hline
$-$3/4 & 10. & 257(1)& 519(3) & 1318(8) & 2671(15) &
       5372(31)& 10747(64) & \\
\hline
$-$1/2 & 5.  & 322(1)& 716(4) & 2088(12) & 4689(27) &
       10379(61)& 23336(138) & \\
$-$1/2 & 10. & 458(2)& 1052(6)& 3092(18)& 6964(41)&
       15736(96)  & 35485(214) & \\
$-$1/2 & 34. & 726(3)& 1859(11)& 6034(39)& 14344(94)&
       32933(222) & 74819(522) & \\
\hline
$-$3/8 &  5. & 430(2)& 1035(6)& 3251(19) & 7746(47) &
       18399(116)  & 43618(270) & \\
$-$3/8& 13. & 658(3)& 1667(10)& 5513(35)& 13613(89)&
       32933(222) & 79261(456) & \\
\hline
$-$1/4 & 0.7 & 212(1)& 507(3)& 1615(9)& 4001(23)&
       9708(57) & 23969(140) & \\
$-$1/4 & 4.7 & 552(2)& 1405(8)& 4842(30)& 12376(81)&
       31353(214) & 78563(550) & \\
\hline
 0 & 1.0 & 458(2)& 1259(7)& 4842(30)& 13613(89)&
       37872(266) & 106280(806) & \\
\hline
& & \multicolumn{7}{||c||}{$\lambda=0.5$}\\ \hline
$-$1/2   & 22. & 529(2)& 1176(7) & 3174(20)& 6640(42)&
       13593(49) & 27748(167) & \\
\hline
$-$1/4 & 1.0 & 193(1)& 428(2)& 1236(8)& 2778(17)&
       6163(38) & 13877(86) & \\
$-$1/4 & 3.3 & 354(1)& 829(5)& 2525(16)& 5912(37)&
       13593(49) & 31605(204) & 72165(472)\\
$-$1/4 & 6.0 & 488(2)& 1168(7)& 3690(22)& 8784(58)&
       20578(138) & 48194(330) & 112036(776)\\
\hline
 0   & 0.5 & 238(1)& 592(3)& 2026(12)& 5198(32)&
       13593(49) & 35624(232) & \\
 0   & 1.0 & 342(1)& 888(5)& 3174(20)& 8390(55)&
       22322(151) & 59257(410) & \\
 0   & 2.0 & 502(2)& 1339(6)& 4925(23)& 13215(64)&
       35925(190) & 96580(528) & 262693(1550) \\
 0   & 10. & 767(2)& 2133(12)& 8320(56)& 23177(170)&
       64877(506) & 182090(1524) & 506839(4560)\\
\hline
& & \multicolumn{7}{||c||}{$\lambda=0.9$}\\ \hline
$-$1/10 & 0.2 & 115.4(0.6) & 237.1(1.6)& 608(4)&1228(9) &
            2528(20)& 5134(36)  & 10465(74) \\
$-$1/10 & 0.5 & 138.4(0.6) & 289.7(2.0)& 766(5)&1594(11) &
            3242(22)& 6743(46)  & 13839(96) \\
$-$1/10 & 1.0 & 174.1(0.8) & 373.2(2.5)& 1006(7)&2119(14) &
            4419(30)& 9287(65)  & 19391(135) \\
\hline \hline
\end{tabular}
\caption{Values of $R^2_e$ for $\delta\leq 0$. All error bars are two standard
         deviations.
        }
\label{table1}
\end{table}
\clearpage

\begin{table}
 \protect\footnotesize
\hspace*{-2cm}\begin{tabular}{|cc|ccccccc|}
\hline\hline
$\delta$ & g & 100 & 200 & 500 & 1000 & 2000 & 4000 & 8000\\ \hline
& & \multicolumn{7}{||c||}{$\lambda=0.6$}\\ \hline
 0   & 1.0 & 304(1)& 754(5)& 2582(16)& 6548(42)&
       16710(113) & 42671(303) & \\
 0   & 2.0 & 448(2)&1156(7)& 4080(27)& 10614(74) &
       27816(205)& 71843(552) & \\
 0   & 4.0 & 620(3)& 1653(10)& 6075(42)& 16173(118)&
       43199(344)& 113130(980)& \\
 0   & 5.0 & 664(3)& 1800(11) & 6672(48) & 17927(136) &
       48107(390)& 129357(1110)& \\
\hline
& & \multicolumn{7}{||c||}{$\lambda=0.75$}\\ \hline
 0   & 0.5 & 182(1)& 412(3)& 1236(8)& 2816(19)&
       6571(44)& 15224(100) & \\
 0   & 1.0 & 253(1)& 599(4)& 1865(12)& 4473(30)&
       10599(72)& 24984(174) & \\
 0   & 5.0 & 593(3)& 1523(10)& 5253(39)& 13515(106) &
       33763(288)& 84913(750) & \\
\hline
1/10 & 0.5 & 224(1) & 541(3) & 1784(12) & 4479(30) & 11058(76)
      & 28044(202) & \\
1/10 & 1.0 & 326(1) & 829(5) & 2845(19) & 7275(52) & 18615(178)
      & 47802(370) & \\
1/10 & 2.0 & 481(2) & 1262(8) & 4503(32) & 11705(90) & 30741(250)
      & 79394(680) & \\
\hline
1/4  & 1.0 & 479(2) & 1330(9) & 5117(38) & 14303(112) & 40132(342)
     & 111634(1042) & \\
1/4  & 2.0 & 642(2) & 1773(12) & 6842(50) & 19183(148) & 53193(450)
     & 147806(1320) & \\
\hline
& & \multicolumn{7}{||c||}{$\lambda=0.9$}\\ \hline
0   & 0.3 & 136(1) & 290(2)& 789(5)& 1678(11) &
            3605(25)& 7703(54) & 16455(114)\\
0   & 0.5 & 160(1) & 348(2)& 967(7)& 2127(15) &
            4579(32)& 10022(68) & 21777 (145) \\
0   & 1.0 & 214(1) & 479(3)& 1400(9)& 3131(22) &
             6943(51)& 15525(110) & 34318 (244)\\
\hline
2/10& 0.2 & 159(1) & 367(2) & 1141(8) & 2727(19) &
            6593(46) & 16116(116)& 39180(288) \\
2/10& 0.5 & 240(1) & 596(4) & 1990(14) & 5003(37) &
            12662(96) & 31804(252) & 80928 (675) \\
2/10& 1.0 & 353(2) & 911(6) & 3169(24) & 8116(64) &
            21001(176) & 53719(482) & 135749 (1290)\\
\hline
& & \multicolumn{7}{||c||}{$\lambda=1$}\\ \hline
0    &  0.1 & 109.2(0.5)&  224.0(1.6)&  571(4) & 1167(8) &
                     2393(17)&  4861(34) & 9866(70) \\
0    &  0.5 & 148.5(0.7)&  318.8(2.2)&  857(6) & 1810(12) &
                     3824(27)&  8104(56) & 17021(122) \\
0    &  1.0 & 192.8(0.9)&  426.3(2.8)& 1190(8) & 2569(18) &
                     5541(39)& 11879(84) & 25536(186)  \\
\hline
1/3  &  0.2  & 187.7(0.9)  & 462(3)  & 1563(11) & 4016(30) &
        10389(80) & 26944(220) &  70143(616)\\
1/3  &  0.5  & 299.3(1.4)  & 784(5)  & 2829(22) & 7512(62) &
        19567(172) & 51537(478) & 133100(1334)\\
1/3  &  1.00 & 438(2)  & 1176(8)  & 4245(34)    & 11248(96) &
        29304(277) & 77797(771) & 203511(2172)\\
1/3  &  2.15 & 600(3)& 1621(11)& 5906(46)& 15923(133) &
        42261(385) & 112434(1100)& \\
\hline
1/2  &  1.41 & 686(7)& 1907(18)& 7389(71)& 20805(210)&55614(492)&
               154556(1444)& 435115(4298) \\
1/2  &  1.79 & 729(8)& 2038(25)& 7901(70)& 22182(208)&58609(504)&
               163330(1532)& 459835(4484) \\
1/2  &  2.83 & 706(8)& 2000(23)& 7897(70)& 22483(208)& & & \\
\hline
2/3  &  1.00 & 655(6)& 1820(12)& 7091(86)& 19756(185)& & & \\
\hline \hline
\end{tabular}
\caption{Values of $R^2_e$ for $\delta\geq 0$. All error bars are two standard
         deviations.
        }
\label{table3}
\end{table}
\clearpage

\begin{table}
 \protect\footnotesize
\hspace*{-35pt}\begin{tabular}{|cc|llllll|l|}
\hline\hline
$\delta$ & g & 100 & 200 & 500 & 1000 & 2000 & 4000 & $\nu_{th}$\\ \hline
& & \multicolumn{6}{||c||}{$\lambda=0$}&\\ \hline
$-$1 & 10.   & 0.5003(8)& 0.5000(12)& 0.4985(19)& 0.4999(30)& 0.4953(59) &&
0.5000\\
\hline
$-$3/4 & 2.0 & 0.5505(8)& 0.5532(12)& 0.5534(18)& 0.5538(29)& 0.5556(58)
 & & 0.5625 \\
$-$3/4 & 5.0 & 0.5588(8)& 0.5585(12)& 0.5584(18)& 0.5578(29)& 0.5575(56) &&\\
$-$3/4 & 8.0 & 0.5617(8)& 0.5616(12)& 0.5621(18)& 0.5638(28)& 0.5637(57) &&\\
\hline
$-$1/2 & 1.0 & 0.6141(8)& 0.6178(12)& 0.6199(18)& 0.6179(28)& 0.6210(57) &&
0.6250\\
$-$1/2 & 3.0 & 0.6246(8)& 0.6255(12)& 0.6253(19)& 0.6269(30)& 0.6270(59) &&\\
$-$1/2 & 5.0 & 0.6290(8)& 0.6284(12)& 0.6272(19)& 0.6232(30)& 0.6174(62) &&\\
\hline
$-$1/4 & 1.0 & 0.6869(7)& 0.6884(10)& 0.6895(14)& 0.6894(20)&
               0.6886(32)&0.6854(67)&0.6825\\
$-$1/4 & 2.0 & 0.6902(9)& 0.6901(13)& 0.6894(20)&
               0.6915(33)& 0.6942(68)&& \\
$-$1/4 & 4.0 & 0.6969(9)& 0.6952(14)& 0.6945(22)&
               0.6946(35)& 0.6948(72)&& \\ \hline
& & \multicolumn{6}{||c||}{$\lambda=0.25$}&\\ \hline
$-$3/4 & 10. & 0.5068(8)& 0.5062(12)& 0.5046(19)& 0.5022(30)& 0.5002(60)&&
0.5000\\
\hline
$-$1/2 & 5.0 & 0.5807(8)& 0.5810(12)& 0.5797(19)& 0.5787(30)&
0.5844(60)&&0.5833\\
$-$1/2 & 10. & 0.5897(8)& 0.5872(13)& 0.5869(19)& 0.5873(31)& 0.5866(62)&& \\
$-$1/2 & 34. & 0.6329(9)& 0.6177(14)& 0.6051(22)& 0.5958(35)& 0.5919(70)&& \\
\hline
$-$3/8 & 5.0 & 0.6263(8)& 0.6246(13)& 0.6243(20)& 0.6233(31)& 0.6226(64)&&
0.6250\\
$-$3/8 & 13. & 0.6513(9)& 0.6453(14)& 0.6408(21)& 0.6354(35)& 0.6335(70)&&\\
\hline
$-$1/4 & 0.7 & 0.6400(8)& 0.6439(12)& 0.6476(19)& 0.6457(30)&
               0.6520(60)&&0.6666\\
$-$1/4 & 4.7 & 0.6732(9)& 0.6723(14)& 0.6703(21)&
0.6667(34)& 0.6626(71)&& \\ \hline
 0 & 1.0 & 0.7378(9)& 0.7404(14)& 0.7423(22)& 0.7410(36)&
                    0.7443(75) && 0.7500\\
\hline
& & \multicolumn{6}{||c||}{$\lambda=0.5$}& \\ \hline
$-$1/2 & 22. & 0.5382(8)& 0.5277(12)& 0.5212(20)& 0.5158(32)& 0.5147(53)&&
           0.5000\\ \hline
$-$1/4 &1.0 & 0.5794(9)& 0.5805(13)& 0.5808(20)& 0.5801(31)& 0.5855(63)&&
0.6250\\
$-$1/4 & 3.3 & 0.6076(7)& 0.6061(10)& 0.6047(15)& 0.6026(20)& 0.6030(26)&
     0.5956(66) & \\
$-$1/4 & 6.0 & 0.6215(7)& 0.6187(11)& 0.6153(15)& 0.6124(22)& 0.6112(35)&
     0.6085(70) & \\
\hline
 0   & 0.5 & 0.6774(8)& 0.6835(12)& 0.6893(19)& 0.6942(33)& 0.6950(54)&&
0.7500\\
 0   & 1.0 & 0.6980(9)& 0.7012(14)& 0.7039(21)& 0.7051(34)& 0.7042(70)&& \\
 0   & 2.0 & 0.7139(6)& 0.7153(8) & 0.7171(12)& 0.7183(17)& 0.7174(29)&
             0.7218(59) & \\
 0   &10.0 & 0.7410(8)& 0.7417(12) & 0.7415(18)& 0.7422(26)& 0.7416(43)&
             0.7384(89) & \\
\hline
& & \multicolumn{6}{||c||}{$\lambda=0.9$}& \\ \hline
$-$1/10 & 0.2& 0.5142(8)& 0.5135(12)& 0.5138(16)&
           0.5149(23)& 0.5124(38)& 0.5137(72)& 0.5000\\
$-$1/10 & 0.5& 0.5258(8)& 0.5236(11)& 0.5216(16)&
           0.5206(22)& 0.5235(35)& 0.5186(70)& \\
$-$1/10 & 1.0 & 0.5381(8)& 0.5352(11)& 0.5334(16)&
           0.5326(22)& 0.5334(35)& 0.5311(72)& \\
\hline \hline
\end{tabular}
\caption{Values of $\nu_{eff}$   for $\delta\leq 0$, for various values of
$N_{cut}$.
        }
\label{table2}
\end{table}
\clearpage

\begin{table}
 \protect\footnotesize
\hspace*{-35pt}\begin{tabular}{|cc|llllll|l|}
\hline\hline
$\delta$ & g & 100 & 200 & 500 & 1000 & 2000 & 4000 & $\nu_{th}$\\ \hline
& & \multicolumn{6}{||c||}{$\lambda=0.60$}& \\ \hline
 0   & 1 & 0.6698(9) & 0.6733(14)& 0.6745(22)& 0.6761(35)& 0.6763(70) &&
0.7000\\
 0   & 2 & 0.6886(9) & 0.6898(15)& 0.6903(23)& 0.6899(37)& 0.6845(77) &&\\
 0   & 4 & 0.7072(10)& 0.7063(16)& 0.7041(25)& 0.7020(41)& 0.6945(85) &&\\
 0   & 5 & 0.7147(10)& 0.7134(16)& 0.7128(25)& 0.7128(41)& 0.7135(85) &&\\
\hline
& & \multicolumn{6}{||c||}{$\lambda=0.75$}&\\ \hline
 0   & 0.5 & 0.5996(9) & 0.6023(14)& 0.6045(21)& 0.6086(34)&
0.6061(68)&&0.6250\\
 0   & 1.0 & 0.6229(9) & 0.6234(14)& 0.6239(22)& 0.6204(35)& 0.6186(70)&&\\
 0   & 5.0 & 0.6745(10)& 0.6718(16)& 0.6687(26)& 0.6628(43)& 0.6653(89)&&\\
\hline
1/10 & 0.5 & 0.6537(9) & 0.6582(14)& 0.6612(22)& 0.6614(35)&
0.6713(72)&&0.6750\\
1/10 & 1.0 & 0.6754(10)& 0.6766(15)& 0.6783(24)&
0.6790(38)& 0.6803(77)&&\\ 1/10 & 2.0 & 0.6927(10)& 0.6918(16)& 0.6908(25)&
0.6907(41)& 0.6844(85)&&\\ \hline
1/4  & 1.0 & 0.7385(10)& 0.7397(17)& 0.7416(27)& 0.7413(44)&
0.7380(91)&&0.7500\\
1/4  & 2.0 & 0.7375(10)& 0.7386(16)& 0.7387(26)&
0.7364(42)& 0.7372(89)&& \\  \hline
& & \multicolumn{6}{||c||}{$\lambda=0.90$}&\\ \hline
 0   & 0.3 & 0.5470(8)& 0.5476(11)& 0.5481(15)& 0.5487(22)& 0.5476(35)&
                    0.5475(70) & 0.5500 \\
 0   & 0.5 & 0.5608(8)& 0.5608(11)& 0.5609(16)& 0.5598(23)& 0.5624(36)&
                    0.5598(71) & \\
 0   & 1.0 & 0.5798(8)& 0.5788(11)& 0.5770(16)& 0.5762(23)& 0.5763(36)&
                    0.5722(72) & \\ \hline
2/10 & 0.2 & 0.6271(8)& 0.6336(11)& 0.6382(16)& 0.6411(23)& 0.6428(37)&
                    0.6408(74) & 0.6500\\
2/10 & 0.5 & 0.6631(8)& 0.6656(12)& 0.6679(17)& 0.6687(25)&0.6689(41)&
                    0.6737(83) & \\
2/10 & 1.0 & 0.6803(9)& 0.6793(13)& 0.6787(19)& 0.6778(28)& 0.6733(46)&
                    0.6687(94) &\\  \hline
& & \multicolumn{6}{||c||}{$\lambda=1$}&\\ \hline
0& 0.1& 0.5141(8)& 0.5136(12)& 0.5139(16)& 0.5131(23)& 0.5109(36)& 0.5106(72)&
0.5000\\
0& 0.5& 0.5411(8)& 0.5394(12)& 0.5393(16)& 0.5392(23)& 0.5386(36)&
0.5353(72)&\\
0& 1.0& 0.5579(8)& 0.5546(11)& 0.5529(16)& 0.5520(23)& 0.5511(37)&
0.5521(73)&\\
\hline
1/3& 0.20& 0.6741(8)& 0.6811(13)& 0.6859(18)& 0.6876(26)& 0.6888(42)&
           0.6902(86)&0.6666\\
1/3& 0.50& 0.6972(9)& 0.6969(14)& 0.6949(20)& 0.6921(29)& 0.6918(48)&
           0.6844(99)&\\
1/3& 1.00& 0.7014(10)& 0.6987(14)& 0.6979(21)& 0.6969(31)& 0.6992(51)&
           0.6937(105)&\\
1/3& 2.15& 0.7099(11)& 0.708(18)& 0.7083(28)& 0.7049(46)& 0.7058(97)&&\\
\hline
1/2& 1.41& 0.7348(13)& 0.7341(16)& 0.7326(22)& 0.7320(32)& 0.7418(48)&
        0.7466(98)&0.7500\\
1/2& 1.79& 0.7332(13)& 0.7315(17)& 0.7302(21)& 0.7300(31)& 0.7428(47)&
        0.7467(98)&\\
1/2& 2.83& 0.7514(29)& 0.7518(46)& 0.7547(92)&&&&\\
\hline
2/3  &  1.00 & 0.7402(27) & 0.7410(35) & 0.7391(111) &&& &0.7500\\
\hline \hline
\end{tabular}
\caption{Values of $\nu_{eff}$ for $\delta\geq 0$, for various values of
$N_{cut}$.
        }
\label{table4}
\end{table}

\end{document}